\documentclass[journal=apchd5,manuscript=article, layout=twocolumn]{achemso}
\usepackage[T1]{fontenc}
\usepackage{stix2} 
\usepackage[scaled=0.9]{helvet}  
\usepackage{orcidlink}
\graphicspath{{Figs/}}%
\usepackage[separate-uncertainty=true, multi-part-units=single]{siunitx}
\usepackage{chemformula}
\usepackage{amsmath}
\usepackage{amssymb} 

\usepackage{microtype}
\usepackage{hyperref}
\definecolor{CustomBlue}{HTML}{0f75bd}
\hypersetup{
    colorlinks=true,
    linkcolor=CustomBlue,
    urlcolor=CustomBlue,
    citecolor=CustomBlue
		}
\usepackage[normalem]{ulem}
\usepackage{caption}
\usepackage{titlesec}
\titleformat{\section}[block]{\normalsize\bfseries\color{CustomBlue}}{\normalsize$\blacksquare$}{0.5em}{} 
\titleformat{\subsection}[runin]{\normalsize\bfseries}{}{0.5em}{}[.]
\titlespacing*{\section}{2pt}{*2}{5pt}
\titlespacing*{\subsection}{0pt}{*1}{5pt}
\renewenvironment{acknowledgement}{%
  {\large\bfseries\color{CustomBlue}\normalsize$\blacksquare$\hspace{0.5em}Acknowledgement}%
  \par\vspace{0.5em}%
}{}
\renewenvironment{suppinfo}{%
  {\large\bfseries\color{CustomBlue}\normalsize$\blacksquare$\hspace{0.5em}Supporting Information}%
  \par\vspace{0.5em}%
}{}

\renewenvironment{tocentry}{%
 \onecolumn%
 \vspace{2em}
 {\large\bfseries\color{CustomBlue}\normalsize$\blacksquare$\hspace{0.5em}TOC Graphic}%
\par\vspace{0.5em}%
\begin{center}%
}{%
\end{center}%
}

\usepackage{fancyhdr} 
\author{Sanatan Halder \orcidlink{0009-0002-2457-0449}}
\email{sanatanh@iitk.ac.in}
\affiliation[IITK]{Department of Physics, Indian Institute of Technology Kanpur, Kanpur - 208016, India}

\author{Manas Khan \orcidlink{0000-0001-6446-3205}}
\email{mkhan@iitk.ac.in}
\affiliation[IITK]{Department of Physics, Indian Institute of Technology Kanpur, Kanpur - 208016, India}

\title{Dynamically stable optical trapping of thermophoretically active Janus colloids}


\begin{document}

\begin{abstract} 
  The ability to optically trap and manipulate artificial microswimmers such as active Janus particles (JPs) provides a breakthrough in active matter research and applications. However, it presents significant challenges because of the asymmetry in the optical properties of JPs and remains incomprehensible. Illustrating the interplay between optical and thermophoretic forces, we demonstrate dynamically stable optical trapping of Pt-silica JPs, where the force-balanced position evolves spontaneously within a localized volume around the focal point and in a vertically shifted annular confinement at low and high laser powers, respectively. Intriguingly, the orientational and orbital dynamics of JP remain strongly coupled in the delocalized confinement. Furthermore, we demonstrate simultaneous optical trapping of multiple JPs. This first report on thermophoresis of Pt-silica JPs and localized-to-delocalized crossover in the position distributions of an optically trapped active JP, verifying theoretical predictions, advances our understanding on confined active matter and their experimental realizations.
\end{abstract}

\section{Introduction}

The pioneering discovery of optical tweezers \cite{ashkinAccelerationTrappingParticles1970, ashkinObservationSinglebeamGradient1986, ashkinOpticalTrappingManipulation1987a, ashkinOpticalTrappingManipulation1987, ashkinForcesSinglebeamGradient1992, ashkinOpticalTrappingManipulation1997} has revolutionized micromanipulation across various fields of science and engineering by enabling the trapping, transport, and manipulation of micron and submicron particles, including living cells, transparent microspheres, birefringent, and metal particles, with precise control \cite{ashkinAccelerationTrappingParticles1970, ashkinObservationSinglebeamGradient1986, ashkinOpticalTrappingManipulation1987a, ashkinOpticalTrappingManipulation1987, svobodaBiologicalApplicationsOptical1994, satoOpticalTrappingMicroscopic1994, svobodaOpticalTrappingMetallic1994, ashkinOpticalTrappingManipulation1997, Sheetz1998, Dholakia2008, Gennerich2017, ashkinOpticalTweezersTheir2018, zhuOpticalTweezersStudies2020, konyshevModelSystemsOptical2021, Volpe2023, halderOpticalMicromanipulationSoft2024a}. The recent surge in fundamental studies and applications of self-propelled microparticles \cite{Ramaswamy2010, Romanczuk2012, Marchetti2013, Cates2015, Bechinger2016, Fodor2018} has motivated efforts to optically trap and manipulate these active particles. Synthetic microswimmers are most conveniently realized using half-metal-coated Janus particles (JPs), which harness phoretic self-propulsion from a local chemical or temperature gradient generated by the metal cap \cite{howseSelfMotileColloidalParticles2007, jiangActiveMotionJanus2010, Buttinoni2012, Wilson2012, simoncelliCombinedOpticalChemical2016,  ilicExploitingOpticalAsymmetry2016, tkachenkoEvanescentFieldTrapping2023, brontecirizaOpticallyDrivenJanus2023}. The distinct optical and photonic properties of the two hemispheres complicate the interaction of the JPs with the trapping laser field and generate a local temperature difference, inducing a thermophoretic force that increases monotonically with the laser power \cite{jiangActiveMotionJanus2010, simoncelliCombinedOpticalChemical2016,  ilicExploitingOpticalAsymmetry2016, tkachenkoEvanescentFieldTrapping2023, brontecirizaOpticallyDrivenJanus2023}.

These complex optical and thermophoretic forces have been utilized in recent studies to achieve intriguing dynamics of JPs, ranging from spinning \cite{merktCappedColloidsLightmills2006, zongOpticallyDrivenBistable2015}, linear transport \cite{nedevOpticallyControlledMicroscale2015, ilicExploitingOpticalAsymmetry2016, liuSelfpropelledRoundtripMotion2016,  tkachenkoEvanescentFieldTrapping2023}, and orbital motion \cite{liuRayopticsModelOptical2015, zongOpticallyDrivenBistable2015, schmidtMicroscopicEnginePowered2018, paulOptothermalEvolutionActive2022, brontecirizaOpticallyDrivenJanus2023} in optical traps. However, a comprehensive understanding of the stable optical trapping of JPs, favorable conditions, and possible trapping configurations remains inadequate. Furthermore, a predicted crossover in the position distribution of a harmonically bound active Brownian particle (HBABP), from Boltzmann-like localized to a delocalized bimodal, based on the strength of confinement and activity \cite{Pototsky2012, Basu2019, Malakar2020}, has been experimentally verified in similar systems \cite{Takatori2016, schmidtNonequilibriumPropertiesActive2021, Buttinoni2022}, but not with an optically trapped phoretically active JP.

Here, we demonstrate dynamically stable optical trapping of half-Pt-coated silica (Pt-silica) JPs in a linearly polarized optical trap and provide a comprehensive description of the relative strengths and orientations of the optical and thermophoretic forces, which evolve spontaneously with the position-orientation of the JP. A Pt-silica JP remains confined to a local three-dimensional (3D) region near the focal point of the trap at lower laser powers and is pushed to a radial distance where the thermophoretic force is balanced by the optical forces, leading to delocalized confinement in an annulus away from the focal plane at a higher laser power. The effective confinement potentials obtained from the position distributions corroborate our findings. Intriguingly, the stochastic orientational dynamics of the JP remains strongly correlated to its orbital motion, analogous to \textit{spin-orbit coupling}, as the Pt-coated hemisphere continues to point radially inward in the delocalized trapping state. We further demonstrate the stable trapping of multiple JPs at different delocalized regions within the laser field. To the best of our knowledge, this is the first observation of thermophoresis with Pt coating and the stable optical trapping of phoretically active JPs.

\section{Forces on a Pt-silica Janus particle}

\subsection{Optical forces}

\begin{figure*}[t!]
	\centering
	\includegraphics[width=1.0\textwidth]{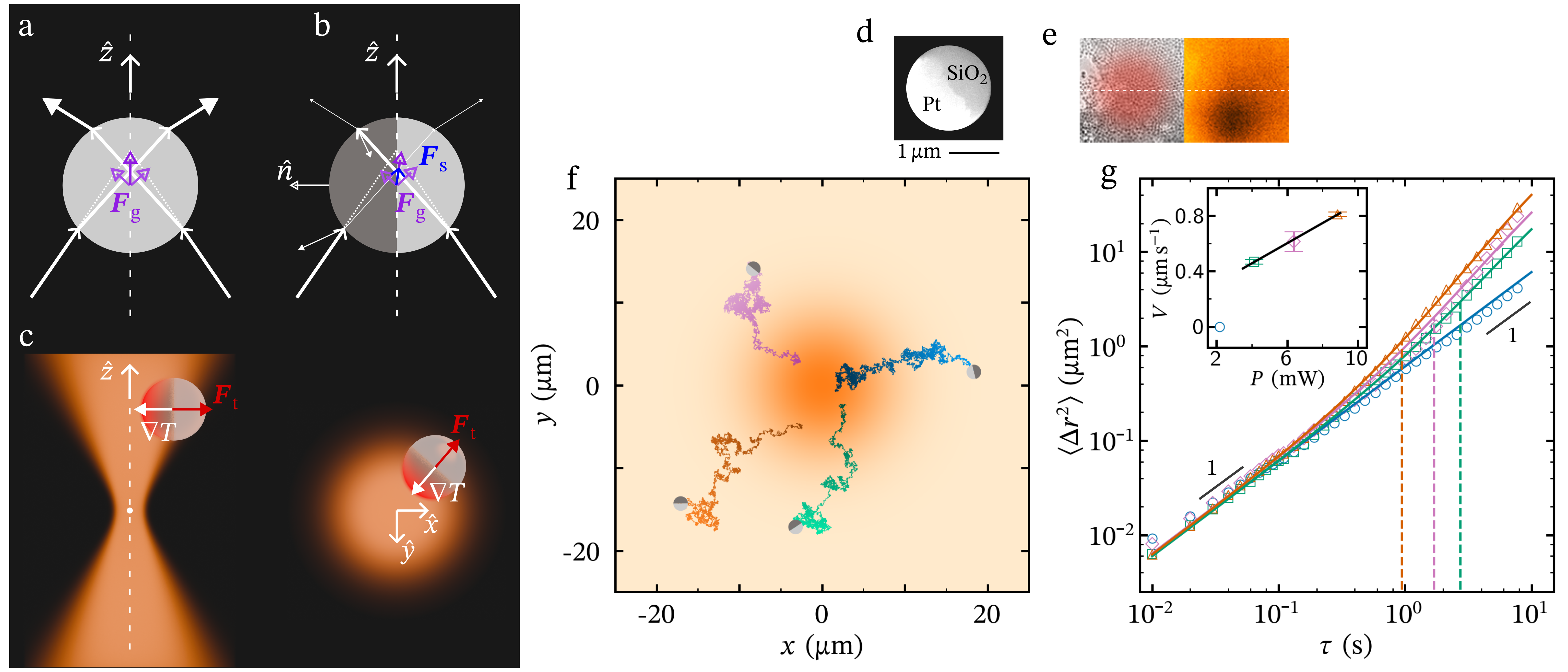}
	\caption{Optical and thermophoretic forces experienced by Pt-silica Janus particles in a laser beam propagating along $\hat{z}$. (a-b) The generation of the gradient force $\mathbfit{F}_{\mathrm{g}}$ and scattering force $\mathbfit{F}_{\mathrm{s}}$ are pictorially explained by showing the path of a pair of typical incoming rays for a particle situated on the beam axis, below the focal point. (a) For a transparent dielectric particle, symmetric components and net $\mathbfit{F}_{\mathrm{g}}$ are exhibited; relatively weak $\mathbfit{F}_{\mathrm{s}}$ is not shown. (b) Axis-asymmetric $\mathbfit{F}_{\mathrm{g}}$ and $\mathbfit{F}_{\mathrm{s}}$ are shown for a Pt-coated (dark grey, along $\hat{n}$) silica Janus particle, whose FESEM image is exhibited in (d). (c) Schematic shows the temperature gradient $\mathbfit{\nabla} T$ (red gradient) generated across the Janus particle and resultant thermophoretic force $\mathbfit{F}_{\mathrm{t}}$. The inset shows the top view. (e) Brightfield micrograph (left) exhibits a monolayer of silica particles, where the bottom half of the surface is Pt-coated, under laser exposure (red overlay). Reduced fluorescence emission of Rhodamine B at the hotter region is shown in orange gradient (right). (f) Four typical trajectories of thermophoretically active Pt-silica Janus colloids (shown at the end points) in a defocused laser beam are shown for \SI{100}{\s} at various laser power $P$ = \SI{2.1}{\mW} (blue), \SI{4.1}{\mW} (green), \SI{6.3}{\mW} (purple), and \SI{8.8}{\mW} (orange). (g) Corresponding MSDs, their fitting with Eq. \ref{eq:msd-abp}, and the crossover times ($\tau_{\mathrm{c}}$) are represented by open symbols, solid curves, and dashed vertical lines of the same color as those of the trajectories, respectively. The linear variation of propulsion speed $V$ with $P$ after a threshold value is shown in the inset.}
	\label{fig:Forces}
\end{figure*}

The optical trap used in this study is a conventional one, realized by tightly focusing a linearly polarized laser beam with a Gaussian intensity profile ($\mathrm{TEM}_{00}$) at $\lambda$ = \SI{1064}{\nm} through a 1.4 NA 60$\times$ objective (Supporting Information). Unlike isotropic dielectric microspheres, for which the interactions and resultant potentials are well understood \cite{ashkinObservationSinglebeamGradient1986, ashkinForcesSinglebeamGradient1992, Sheetz1998, Gennerich2017, halderOpticalMicromanipulationSoft2024a}, the Pt-coated and uncoated hemispheres of a Pt-silica JP interact disparately with the trapping laser beam. The trapping forces can be satisfactorily explained using ray optics in the Mie regime,\textit{ i.e.}, when the particle size (diameter 2$a$) is larger than the working wavelength ($\lambda$) \cite{ashkinForcesSinglebeamGradient1992, Sheetz1998, Gennerich2017, halderOpticalMicromanipulationSoft2024a}. For transparent dielectric particles, the incoming rays are mostly refracted as they pass through, and the corresponding net change in momentum generates a gradient force, $\mathbfit{F}_{\mathrm{g}} \propto \mathbfit{\nabla}I$, where $I$ is the intensity of the laser field. Therefore, $\mathbfit{F}_{\mathrm{g}}$ varies with the position of the particle and creates a restoring force field that vanishes at the focal point (Figure \ref{fig:Forces}(a)), which becomes a time-independent stable trapping position. Weak reflection and nearly zero absorption, if any, result in a small scattering force $\mathbfit{F}_{\mathrm{s}}$, which shifts the stable trapping point marginally along the beam propagation direction $\hat{z}$.

While the uncoated hemisphere of the Pt-silica JP (2$a$ = \SI{1.76}{\um}, Figure \ref{fig:Forces}(d), S2(a)) is almost transparent, the Pt-coating (thickness $\approx$ \SI{5.5}{\nm}) reflects and absorbs a considerable fraction of light with $\approx$ 34\% reflectance and $\approx$ 45\% absorbance at $\lambda$ = \SI{1064}{\nm} \cite{wernerOpticalConstantsInelastic2009, Kulikova2020}, leading to a significant scattering force. Therefore, $\mathbfit{F}_{\mathrm{g}}$, $\mathbfit{F}_{\mathrm{s}}$, and their resultant depend on the position as well as orientation ($\hat{n}$) of the JP, and do not always direct towards the focal point. When $\hat{n}$ is perpendicular to $\hat{z}$, neither $\mathbfit{F}_{\mathrm{g}}$ nor the resultant optical force ($\mathbfit{F}_{\mathrm{g}} + \mathbfit{F}_{\mathrm{s}}$) points along the focal point, as shown in Figure \ref{fig:Forces}(b). Thus, at any particle position, even at the focal point, the net optical force acting on the JP evolves spontaneously with orientational diffusion.

\subsection{Thermophoretic force}

Strong absorption of the laser at the Pt-coated surface makes it hotter, generating a temperature gradient across the Janus colloid (Figure \ref{fig:Forces}(c, e)), and thus inducing a thermophoretic force $\mathbfit{F}_{\mathrm{t}}$, which propels the particle in the medium of deionized water, with velocity $\mathbfit{V} = - D_{\mathrm{th}} \mathbfit{\nabla}T$, where $D_{\mathrm{th}}$ is the thermodiffusion coefficient \cite{duhrWhyMoleculesMove2006, jiangActiveMotionJanus2010} (Figure S2(b)). Previous studies on the thermophoretic properties of JPs have been performed using Au coatings \cite{jiangActiveMotionJanus2010, nedevOpticallyControlledMicroscale2015, simoncelliCombinedOpticalChemical2016, brontecirizaOpticallyDrivenJanus2023, tkachenkoEvanescentFieldTrapping2023}. Hence, we provide a thorough validation and characterization of thermophoresis of Pt-silica JPs under laser exposure.

We validated the increase in the local temperature at the Pt-coated surface under laser exposure by fluorescent thermometry on a monolayer of silica particles, where half of the top surface was coated with Pt. The enhanced temperature at the Pt-coated side was verified by the decreased fluorescence emission of Rhodamine B in that region \cite{Ross2001, Natrajan2008, Zhou2019} (Figure \ref{fig:Forces}(e), Supporting Information).

To characterize $\mathbfit{F}_{\mathrm{t}}$ at various laser powers, we measured the corresponding propulsion speeds $V$ in the diverging part of the laser beam to minimize the effects of $\mathbfit{F}_{\mathrm{g}}$ and $\mathbfit{F}_{\mathrm{s}}$. Owing to the radially falling intensity of the laser field, the thermophoretic propulsion speed decreased as the JP moved away from the center (Figure \ref{fig:Forces}(f) and S2(d, e), Video S1). The radially outward dynamics with the Pt-coated hemisphere ($\hat{n}$) pointing mostly along $- \hat{r}$ indicates $D_{\mathrm{th}}$ is positive for the Pt-silica microspheres. Furthermore, this orientation induces a higher $\mathbfit{\nabla}T$, and consequently generates a stronger thermophoretic effect, resulting in a higher propulsion speed $V$. Therefore, unlike diffusiophoretic activity \cite{howseSelfMotileColloidalParticles2007, Wilson2012}, these active dynamics are anisotropic and exhibit \textit{negative phototaxis}-like motion \cite{GomezSolano2017}. A few typical trajectories and corresponding MSDs at varying laser powers $P$ are shown in Figure \ref{fig:Forces}(f) and (g), respectively. We determined the average value of $V$ from the time-averaged mean square displacement (MSD) \cite{baileyFittingActiveBrownian2022}, which was calculated from the complete trajectory of the Pt-silica JP as it traversed radially outward and was driven by a progressively decreasing thermophoretic force (Figure S2). $\mathbfit{F}_{\mathrm{t}}$ can be obtained from $\mathbfit{V}$ as $\mathbfit{F}_{\mathrm{t}} = 6 \pi \eta a \mathbfit{V}$, where $\eta$ is medium viscosity.

The thermophoretically active dynamics of Pt-silica JP, considering a flat laser field, can be modeled as that of an active Brownian particle (ABP), and is represented by the following Langevin equations: \cite{Bechinger2016,Basu2019,halder2025interplay}
\begin{equation}
	\begin{split}
\dot{x}(t) & = \xi_x(t) + V\cos\phi(t), \\
\dot{y}(t) & = \xi_y(t) + V\sin\phi(t), \  \text{and} \\
\dot{\phi}(t) & = \xi_{\phi}(t),
	\label{LVEqFree}
	\end{split}
\end{equation}
where $\xi_x (t)$, $\xi_y (t)$, and $\xi_{\phi}(t)$ are the random translational and orientational velocity noise, respectively. The corresponding MSD is given by,
\begin{equation}
  \left\langle \Delta r^2(\tau)\right\rangle = 4D_{\mathrm{T}}\tau + 2V^2{\tau^2_{\mathrm{R}}}\left( \tau/\tau_{\mathrm{R}} + e^{-\tau/\tau_{\mathrm{R}}} -1 \right),
	\label{eq:msd-abp}
\end{equation}
where $\tau_{\mathrm{R}}=1/D_{\mathrm{R}}$ is persistent time over which the autocorrelation of propulsion direction decays, and $D_{\mathrm{T}}$, $D_{\mathrm{R}}$ are the translational and orientational diffusion coefficients, respectively. The experimentally observed MSDs were fitted using Eq. \ref{eq:msd-abp}, to obtain the average values of $V$ and $\tau_{\mathrm{R}}$ at each laser power \cite{baileyFittingActiveBrownian2022}. A linear variation in $V$ with $P$ after a threshold value was observed (Figure \ref{fig:Forces}(g) inset). Furthermore, the MSDs exhibit a crossover from diffusive to active dynamics, with the exponent changing from 1 to 2, at a characteristic crossover time $\tau_{\mathrm{c}}$ (Figure \ref{fig:Forces}(g)), which are in full agreement with respective theoretically predicted values $4D_{\mathrm{T}}/V^2$ (Eq. \ref{eq:msd-abp}). Additionally, because of the Gaussian laser intensity profile, the magnitude of $\mathbfit{F}_{\mathrm{t}}$, and consequently, $V$, decreases with radial distance (Supporting Information). The observed thermophoretic propulsion properties of Pt-silica JP are similar to those of Au-coated Janus particles \cite{jiangActiveMotionJanus2010, nedevOpticallyControlledMicroscale2015, simoncelliCombinedOpticalChemical2016, brontecirizaOpticallyDrivenJanus2023, tkachenkoEvanescentFieldTrapping2023}.

\section{Dynamically stable optical trapping}

Unlike the case of an isotropic dielectric particle, the forces experienced by a metal-coated JP in a focused laser field, $\mathbfit{F}_{\mathrm{g}}$, $\mathbfit{F}_{\mathrm{s}}$, and $\mathbfit{F}_{\mathrm{t}}$, are all time-varying as they change with the orientation or the particle ($\hat{n} (t)$), which evolves spontaneously because of orientational diffusion. This temporal variation is stronger for smaller particles, as $D_{\mathrm{R}}$ varies by $1/a^3$. Therefore, there is no time-independent stable trapping position where a JP can remain in force-equilibrium. In contrast, there are multiple spatially distributed instantaneous stable points given by $\mathbfit{F}_{\mathrm{g}} (t) + \mathbfit{F}_{\mathrm{s}} (t) + \mathbfit{F}_{\mathrm{t}} (t)$ = 0 for time-varying position-orientations of the JP. Thus, a JP is spontaneously driven from one stable position to another with its orientational diffusion while remaining bound in a dynamically stable optical confinement. The spatial distribution of the force-balanced positions shifts with increasing laser power, which changes the relative strengths of the optical and thermophoretic forces.

\begin{figure*}[t!]
	\centering
	\includegraphics[width=1.0\textwidth]{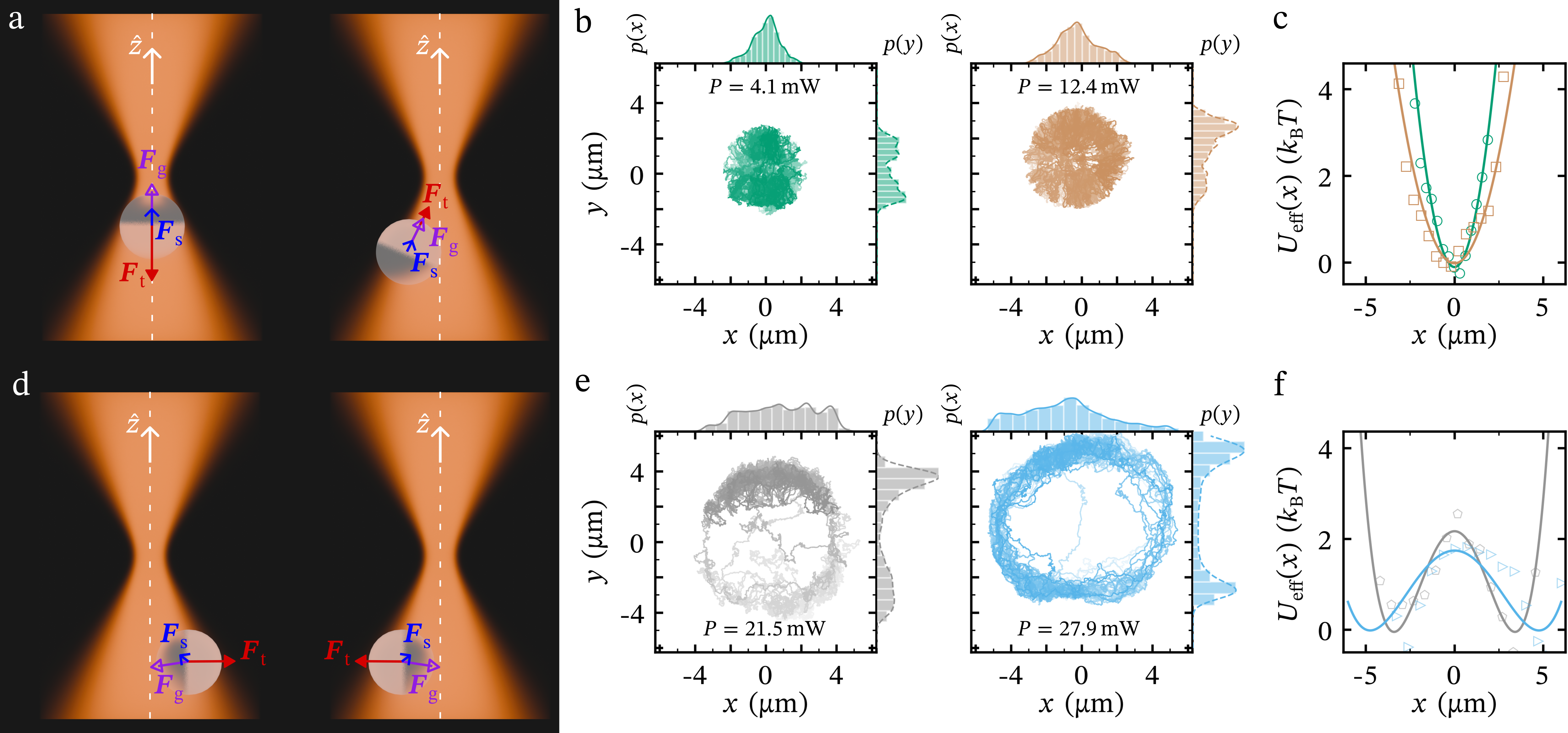}
	\caption{Localized (a – c) and delocalized (d – f) optical confinement of Pt-silica Janus colloids at low and high laser power, respectively. (a, d) Schematics exhibit the directions and relative strengths of the optical forces, $\mathbfit{F}_{\mathrm{g}}$ and $\mathbfit{F}_{\mathrm{s}}$, and thermophoretic force, $\mathbfit{F}_{\mathrm{t}}$, for two typical position-orientations of the Janus colloids in the laser field (orange gradient) propagating along $\hat{z}$, in each case. (b, e) A pair of typical trajectories recorded over \SI{200}{\s} at 500 fps, with corresponding position distributions, $p(x)$ and $p(y)$, are shown, demonstrating (b) localized trapping near the focal point at lower values of $P$,  and (e) delocalized confinement in annular regions at relatively higher values of $P$. (c, f) Effective potentials, $U_{\text{eff}}(x)$, experienced by the Janus particles and their fitting with (c) quadratic and (f) quartic functions are shown with open symbols and solid lines of the same color as those of the trajectories, respectively. }
	\label{fig:DynamicallyStableTrapping}
\end{figure*}

We recorded the trajectories of optically trapped Pt-silica JP over long durations with increasing laser power ($P$) to obtain the spatial distributions of dynamically stable trapping positions. $\mathbfit{F}_{\mathrm{g}}$, $\mathbfit{F}_{\mathrm{s}}$, and $\mathbfit{F}_{\mathrm{t}}$ increase monotonically with $P$, following nontrivial incremental relations, as manifested in the observed crossover from a localized Boltzmann-like to a delocalized bimodal position distributions of an optically trapped JP with increasing $P$. We further derived the effective confining potential $U_{\text{eff}} (x)$ experienced by a thermophoretically active JP using the Boltzmann inversion of steady-state position distributions to describe the variation in the shape of the dynamic optical confinement with increasing laser power. The results are shown in Figure \ref{fig:DynamicallyStableTrapping}.

It is extremely challenging and requires utmost care to record the long trajectories of these thermophoretically active JPs in an optical trap, because the dynamically stable confinement becomes unstable and throws out the JP with even a slight perturbation. Therefore, to obtain even longer trajectories and smoother steady-state position distributions, we simulated the dynamics of an optically trapped thermophoretically active JP, modeling it as an HBABP and following the corresponding Langevin equations: \cite{Pototsky2012, Buttinoni2022, halder2025interplay}
\begin{equation}
	\begin{split}
		\dot{x}(t) & = \xi_x(t) + V\cos\phi(t) - x/\tau_{k}, \\
		\dot{y}(t) & = \xi_y(t) + V\sin\phi(t) - y/\tau_{k}, \  \text{and} \\
		\dot{\phi}(t) & = \xi_{\phi}(t),
		\label{LVEqTrapped}
	\end{split}
\end{equation}
where $\tau_{k} = 6 \pi \eta a / k$ is a characteristic timescale describing the strength of the optical trap, $k$. Our simulations based on this simplified HBABP model did not incorporate the instantaneous position- and orientation-dependent complex interaction of a JP with a tightly focused laser beam. Rather, our goal was to understand the emergent properties at steady-state by simulating the underlying effective nonequilibrium dynamics of this system. Hence, we considered the effect of time-averaged ($\mathbfit{F}_{\mathrm{g}} + \mathbfit{F}_{\mathrm{s}}$) and $\mathbfit{F}_{\mathrm{t}}$ as to provide a radially symmetric restoring force field with force constant $k$ and propel the JP with propulsion speed $V$, respectively. Following our observations, we varied $k$ and $V$ proportionally, as both increase linearly with the laser power $P$ (Figure S3(c), \ref{fig:Forces}(g)), and neglected any variation of $V$ with the instantaneous radial position $r (t)$ of the JP as an optically trapped JP is not expected to explore a wide range of $r$, and the dependence of $V$ on $r$ is considerably weaker than that on $P$ (Supporting Information). The simulation results agreed well with our experimental observations, thereby corroborating our findings. Notably, the position autocorrelations of the optically trapped active JPs showed excellent fit with the analytical prediction, further deepening our understanding of this system.

\subsection{Localized confinement}

At a lower laser power, $\mathbfit{F}_{\mathrm{t}}$ remains comparatively small, while the optical forces push a JP closer to the focal point, as shown in Figure \ref{fig:DynamicallyStableTrapping}(a). Force balance is achieved at various positions near the focal point. For example, all forces align along the optical axis, resulting in a trivial force balance for axis-symmetric orientations ($\hat{n} = \hat{z}$) on the beam axis. Consequently, the JP remains localized near the focal point and spontaneously passes through instantaneous stable positions over a 3D region (Video S2). The recorded trajectories exhibit confined self-similar dynamics with Boltzmann like position distributions peaked at the center of the trap (Figure \ref{fig:DynamicallyStableTrapping}(b), S5(a)). The longer residence time of the particles along the $y$-axis in the experimentally captured trajectories, resulting in a partially distorted $p(y)$, is attributed to the polarization direction of the trapping laser beam (Supporting Information). An increase in the laser power enhances $\mathbfit{F}_{\mathrm{t}}$ more than the optical forces that keep a JP closer to the focal point, resulting in weaker confinement and a wider position distribution. This is evident in $U_{\text{eff}} (x)$, which fits well with the quadratic function $k_{\mathrm{eff}} x^2 / 2$, indicating an effective harmonic potential with stiffness $k_{\mathrm{eff}}$ \cite{Pototsky2012, halder2025interplay} that decreases with increasing laser power (Figure \ref{fig:DynamicallyStableTrapping}(c), S3(b), S5(b)).

\subsection{Delocalized confinement}

At a relatively high laser power, $\mathbfit{F}_{\mathrm{t}}$ becomes stronger and dominates the optical force to push a JP away from the focal point to a radial distance, where $\mathbfit{F}_{\mathrm{t}}$ is balanced by the optical forces $\mathbfit{F}_{\mathrm{g}}$ and $\mathbfit{F}_{\mathrm{s}}$ at a much lower $z$-plane (Figure \ref{fig:DynamicallyStableTrapping}(d)). Therefore, the JP traverses within an annular region with its orientational diffusion, avoiding the center of the trap (Video S3). The recorded trajectories exhibit annular confinement of the JP, which rarely crosses the center. Consequently, the one-dimensional position distributions become bimodal, exhibiting two peaks that shift apart as the diameter of the annulus increases with $P$ (Figure \ref{fig:DynamicallyStableTrapping}(e), S5(d)). Distortions in $p (x)$ in the experimental data arising from the polarization of the trapping laser beam are more pronounced here (Supporting Information). Otherwise, the one-dimensional effective potential, typically denoted by $U_{\text{eff}}(x)$, exhibits two minima and fits well with the quartic function $A x^2 (x^2 - 2 x_{\mathrm{c}}^2)$ in this case, indicating effective harmonic confinement along the radial direction at $r = x_{\mathrm{c}}$, which increases with $P$ (Figure \ref{fig:DynamicallyStableTrapping}(f), S5(e)).

This crossover from localized trapping near the focal point with a Boltzmann-like position distribution to delocalized confinement in an annular region exhibiting bimodal position distribution is similar to the dynamical crossover observed in the HBABP model \cite{Pototsky2012, Basu2019, Malakar2020} and in other types of synthetic ABPs under harmonic confinements \cite{Takatori2016, schmidtNonequilibriumPropertiesActive2021, Buttinoni2022}.

\subsection{Spin-orbit coupling}

\begin{figure}[t!]
	\centering
   \includegraphics	[width=\linewidth]{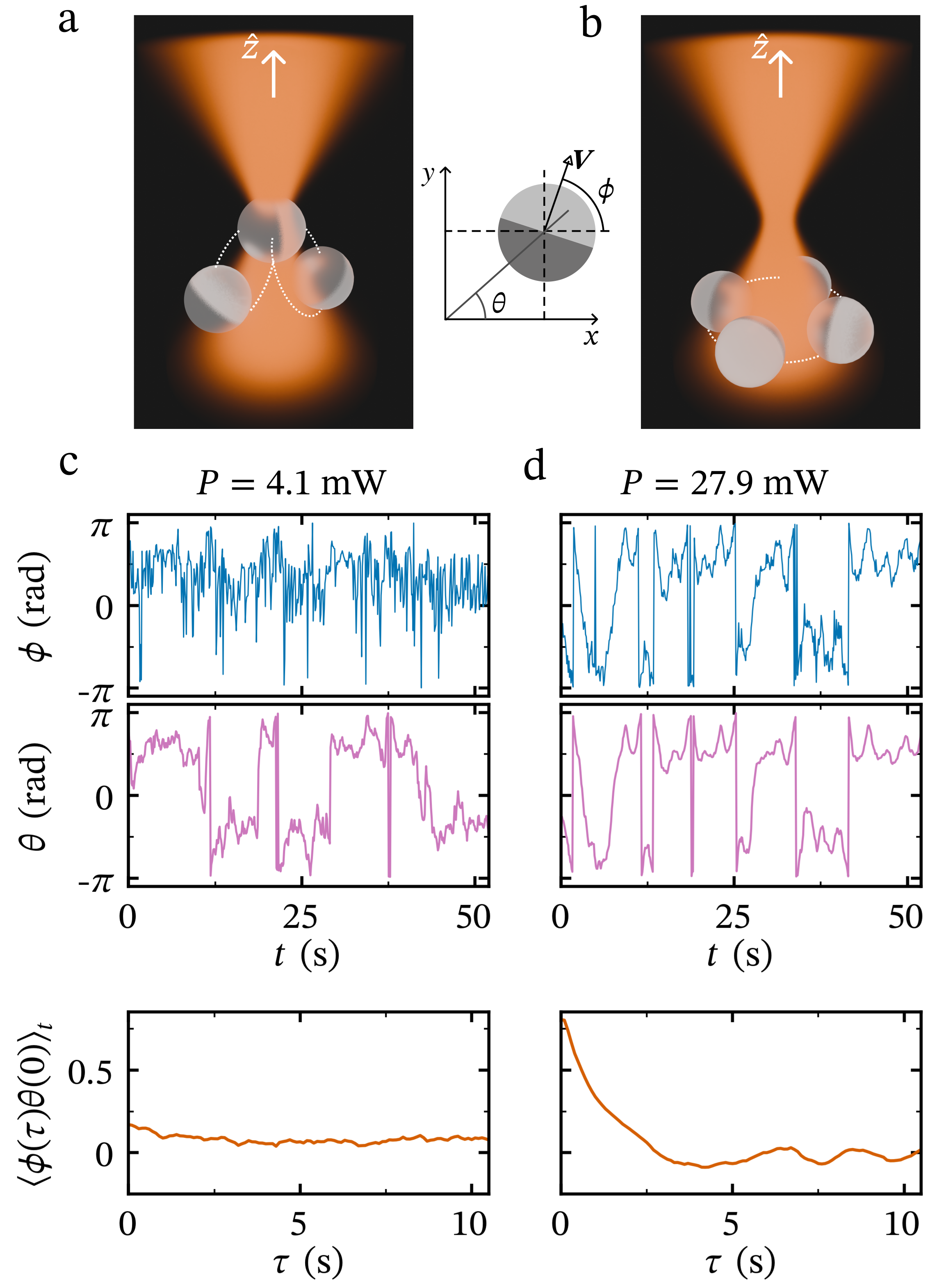}
	\caption{Coupling between the orientational ($\phi(t)$) and orbital ($\theta(t)$) dynamics of an optically trapped Pt-silica Janus colloid at lower (a, c) and higher (b, d) laser powers. (a, b) Schematics exhibit typical stochastic positional and orientational evolution of a Janus particle passing over the force-balanced positions in the optical trap (Video S2, S3). $\phi$ and $\theta$ are pictorially defined in the inset. (c, d) Short parts of the time-series of $\phi(t)$ and $\theta(t)$ and their normalized cross-correlation (time-averaged over \SI{200}{\s}) are shown for the two laser powers.}
	\label{fig:Spin-OrbitCoupling}
\end{figure}

\begin{figure*}[t!]
	\centering
	\includegraphics[width=\linewidth]{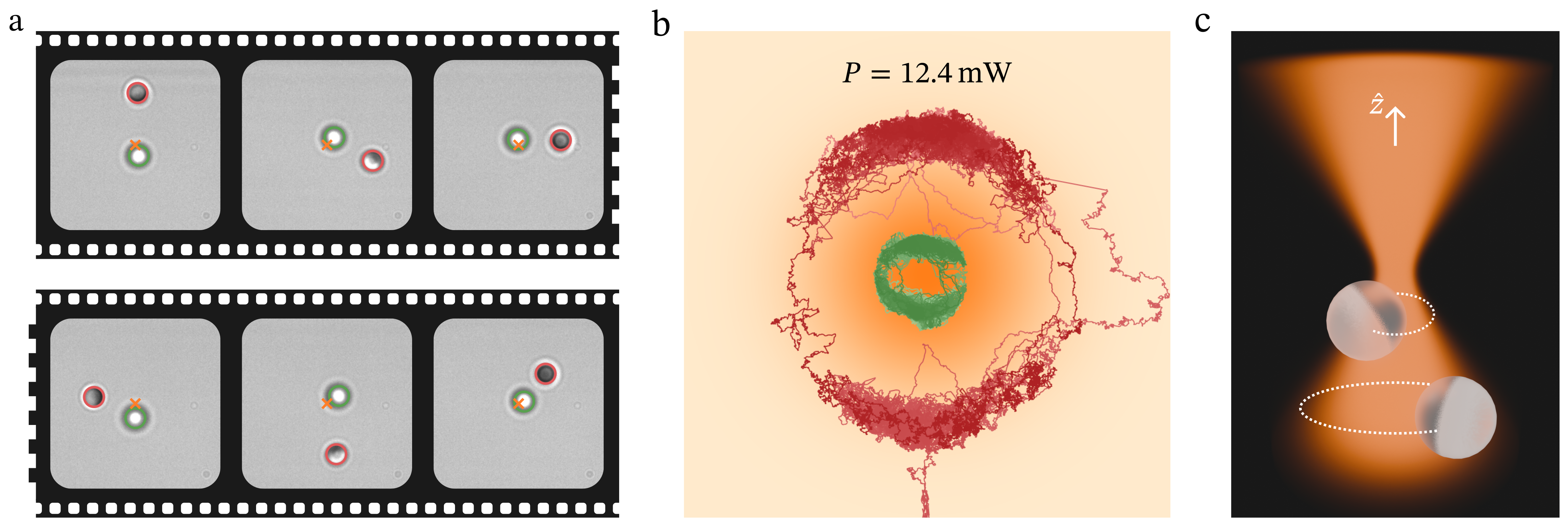}
	\caption{Simultaneous dynamically stable optical trapping of two Pt-silica Janus particles. (a) Snapshots at progressing times from Video S4 show the positions of the Janus particles in reference to the center of the trap (orange cross). The particles are tagged with green and red borders. (b) Annularly confined trajectories of the particles over \SI{200}{\s} are exhibited on an orange gradient representing the laser intensity field. (c) Delocalized confinements, marked by white dotted circles, with typical position-orientations of simultaneously trapped Janus particles are shown schematically.}
	\label{fig:MutipleParticle}
\end{figure*}

Intriguingly, in the case of delocalized confinements in annular regions, a strong correlation between the orientation of the particle $\phi$ and its azimuth $\theta$ (Figure \ref{fig:Spin-OrbitCoupling} (inset)) is observed; this correlation is a stochastic analog of spin-orbit coupling \cite{modinHydrodynamicSpinorbitCoupling2023}. At a lower laser power, a JP remains confined near the focal point, with its orientational and positional dynamics being independent (Figure \ref{fig:Spin-OrbitCoupling}(a), Video S2)). In contrast, when a JP follows annularly confined trajectories, its orientational and positional dynamics become coupled, where $\hat{n}$ always points to $- \hat{r}$, \textit{i.e.}, the propulsion direction $\hat{V}$, which makes an angle $\phi$ with the $x$-axis, continues to be along $\hat{r}$, as $\theta$ evolves stochastically over time (Figure \ref{fig:Spin-OrbitCoupling}(b), Video S3). Short parts of the time series of $\phi(t)$ and $\theta(t)$ and their normalized cross-correlations $\left\langle \phi(t) \theta(t + \tau) \right\rangle_t$ are shown in Figure \ref{fig:Spin-OrbitCoupling}(c, d) for a lower and higher value of $P$, respectively. While there is no correlation at a lower laser power, a strong and slowly decaying correlation emerges along with the delocalization of the optical confinement at a higher $P$. Furthermore, periodic-like variation in $\left\langle \phi(t) \theta(t + \tau) \right\rangle_t$ far beyond the orientational diffusion timescale $\tau_{\mathrm{R}} \approx$ \SI{1.72}{\s} indicates a long-persistent coupling between the orientational and orbital dynamics over their stochastic cycles.

A similar stochastic spin-orbit coupling is also observed in simulated dynamics of an HBABP, with the correlation $\left\langle \phi(t) \theta(t + \tau) \right\rangle_t$ persisting beyond $\tau_{\mathrm{R}}$ (Figure S6). This indicates that the coupling is not caused by optical forces or torques. Rather, it is induced by a faster equilibration of the JP in the harmonic well compared to its orientational diffusion, \textit{i.e.}, when $\tau_k$ is shorter than $\tau_{\mathrm{R}}$, which occurs at higher laser powers (Supporting Information).

\section{Stable optical trapping of multiple JPs}

The condition for localized trapping near the focal point or delocalized confinement in an annular region at a lower $z$-plane is set by the relative strength of the thermophoretic force compared to the optical forces and thus by the laser power. However, small variations in the properties of the JPs, such as the thickness or coverage of the Pt coating, also govern the relative strength among the forces and hence regulate the degree of delocalization of the optical confinement at the same laser power. This allows multiple Janus particles with varied properties to remain dynamically confined in different regions within the 3D laser field. While we observed various multiparticle configurations in the optical trap, a simultaneous stable trapping of two JPs is discussed here. In this case, both the JPs were confined to annular regions with different radii in separate $z$-planes (Video S4). The JP that experiences a smaller $\mathbfit{F}_{\mathrm{t}}$ remains closer to the focal point, \textit{i.e.}, at a nearer $z$-plane in a smaller annulus, compared with the JP that is pushed to a farther radial distance by a stronger $\mathbfit{F}_{\mathrm{t}}$ and stays confined in a larger annulus at a lower $z$-plane. Figure \ref{fig:MutipleParticle} shows snapshots of the progressing time instants representing this observation (Figure \ref{fig:MutipleParticle}(a)), the annularly confined trajectories of the two JPs (Figure \ref{fig:MutipleParticle}(b)), and a schematic of their position-orientations in the laser field (Figure \ref{fig:MutipleParticle}(c)). The asymmetry in the azimuthal position distributions of both JPs is again attributed to the polarization effect (Supporting Information).

\section{Conclusion}

In this study, we demonstrate the dynamically stable optical trapping of Pt-silica JPs, which exhibit thermophoretic activity owing to photonic heating under laser exposure. With a comprehensive description of the optical and thermophoretic forces acting on a JP for its varied position-orientations in a tightly focused laser field, we show that the force-balanced positions shift dynamically with the orientational diffusion. The spatial distributions of the stable trapping positions remain localized around the focal point when the optical forces dominate at a lower laser power and become delocalized in an annulus about the optic axis and away from the focal plane as a stronger thermophoretic force pushes the JP radially outward at a higher laser power. Intriguingly, the stochastic orientational and orbital dynamics of a JP in the delocalized trapping state remain strongly coupled, which is reminiscent of spin-orbit coupling. Furthermore, we demonstrate the simultaneous stable optical trapping of multiple JPs in different regions within the 3D laser field. Inertial and hydrodynamic effects were neglected in this study of the isolated optically trapped dynamics of a JP in the low Reynolds number regime.

Although orbital motions and annular confinements of optically trapped JPs have been previously reported, this is the first observation of localized optical trapping of active JPs and experimental verification of a crossover from localized Boltzmann-like to delocalized bimodal position distribution and thermophoresis of Pt-silica JPs, to the best of our knowledge.  

We believe that our findings provide a breakthrough in the optical micromanipulation of active JPs by deepening our understanding of the complicated interactions between these JPs and a tightly focused laser beam. Precise estimation of the optical and thermophoretic forces acting on a JP in an optical trap, and thus prediction of the spatial distribution of the stable trapping positions at a given laser power through numerical simulations \cite{wangOpticalTrappingJanus2008} and neural network-based calculations \cite{brontecirizaFasterMoreAccurate2023}, provide interesting future directions. Furthermore, our results facilitate and encourage further theoretical and experimental studies on confined and far-from-equilibrium active matter for fundamental research and potential applications, such as microfluidic devices.

\vspace{1em}
\begin{acknowledgement}
Research funding from SERB, Govt. of India through CRG (Grant No. CRG/2020/002723) is gratefully acknowledged. MK thanks Saikat Ghosh for fruitful discussions, which initiated the idea of relating the stochastic dynamics of optically trapped Janus particles to spin-orbit coupling.
\end{acknowledgement}

\begin{description}
  \item[Disclosures] 
  The authors declare no competing interest.

  \item[Data availability]
  All data required to reach the conclusion of this study are presented in the manuscript or Supporting Information.

  \item[Author contributions]
  All the authors contributed to the conception and design of the research. SH conducted the experiments and analyzed the data. SH and MK interpreted data and wrote the manuscript. MK supervised the project.
  
\end{description}

\begin{suppinfo}
	
Supporting Information files include the following: \\
S1: Optical trap setup, S2: Experimental details, S3: Position autocorrelation, S4 Numerical simulation, and S5: Supporting videos\\
{\href{https://drive.google.com/file/d/1OlY6XjiXH8xpVMW_LCGmzbr_DXmnWO6L/view?usp=share_link}{Video S1:}} Thermophoretic active dynamics of a Pt-silica JP in a defocused laser field\\
{\href{https://drive.google.com/file/d/1fx_SvoUPz4wT1VoPAkUc1sLg4SosaSw8/view?usp=share_link}{Video S2:}} Localized optical confinement of a thermophoretically active JP near the focal point\\
{\href{https://drive.google.com/file/d/1-1pv_RiRXyM_syHzbgXgvG93nLSfgYFG/view?usp=share_link}{Video S3:}} Delocalized optical confinement of a thermophoretically active JP in an annularly confined region\\
{\href{https://drive.google.com/file/d/1aCHjMdPPhdLxt2xTzsJW-rwLBcZyX8FJ/view?usp=share_link}{Video S4:}} Coexistent dynamically stable optical trapping of two Pt-silica JPs\\
{\href{https://drive.google.com/file/d/1pBZP3IznVtL25WicY1C9my4H265Scz5_/view?usp=sharing}{Video S5:}} Stochastic spin-orbit coupling from HBABP simulation\\

\end{suppinfo}

\bibliography{references.bib}

\begin{tocentry}
	
	\begin{center}
		\large\textbf{For Table of Contents Use Only}\\
		\vspace{1cm}
		\LARGE\textbf{Dynamically stable optical trapping of thermophoretically active Janus colloids}\\
		\vspace{0.5cm}
			\large{Sanatan Halder and Manas Khan}\\
	\end{center}
	\vspace{2cm}
  \centering
  \includegraphics[width=3.25in, height=1.75in, keepaspectratio]{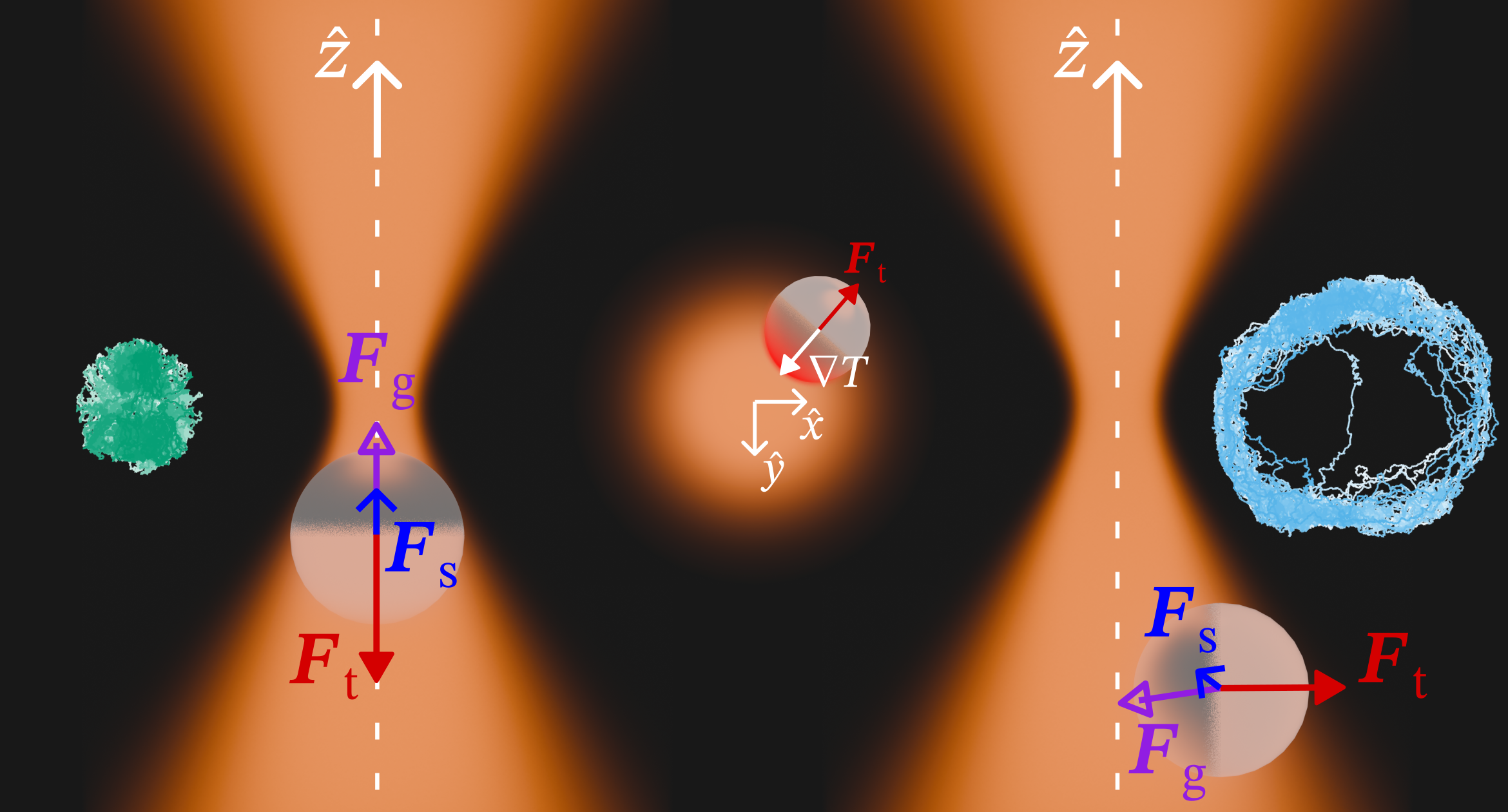}
  \label{fig:toc}
\end{tocentry}
Typical force-balanced configurations and trajectories corresponding to the localized and delocalized optical trapping of a thermophoretically active Janus colloid
\end{document}


\begingroup
\begin{spacing}{1.15} 
	\tableofcontents
\end{spacing}
\endgroup

\section{Optical trap setup}

\begin{figure*}[h]
\centering
\includegraphics[width=0.7\linewidth]{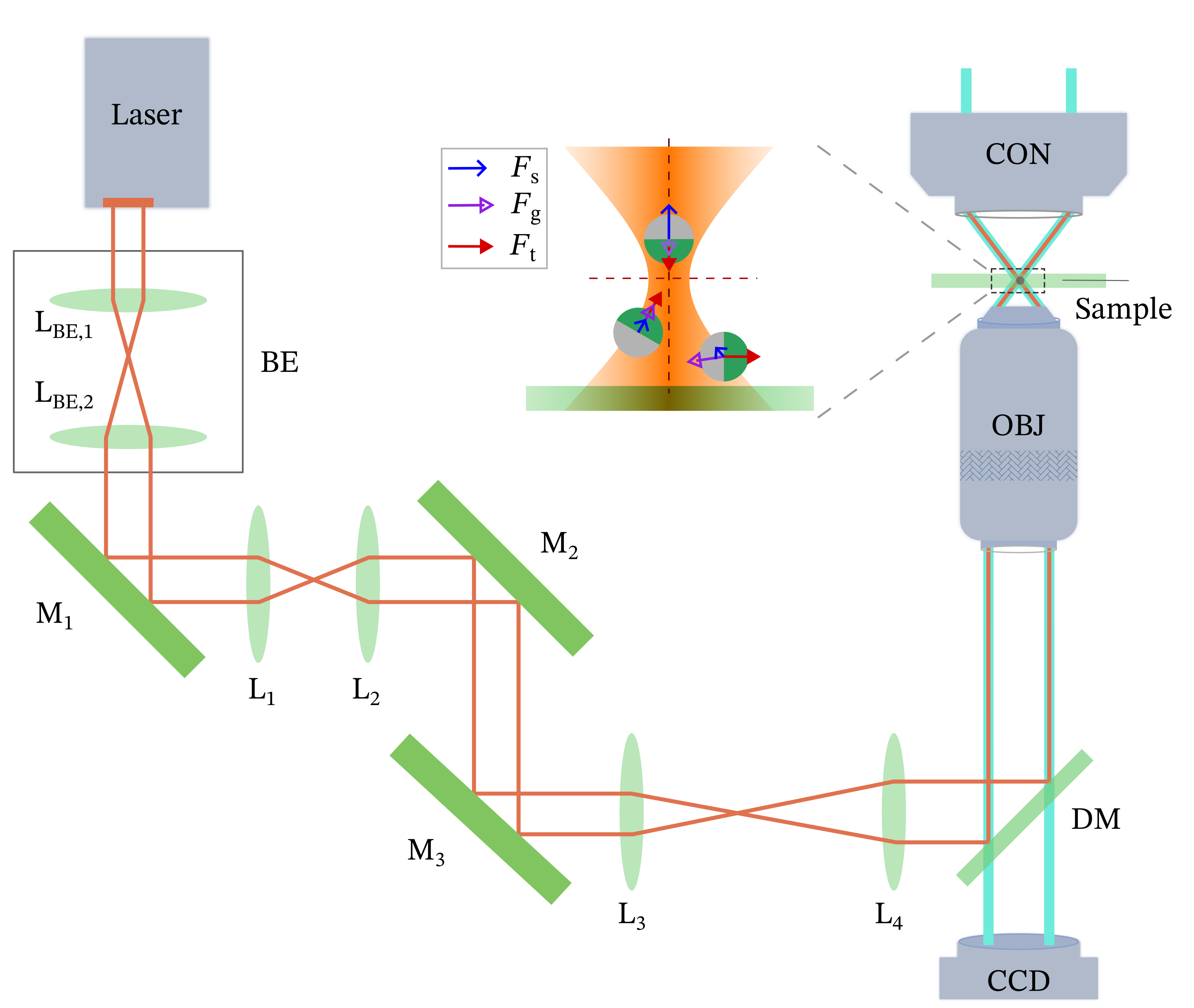}
\caption{Schematic of the optical trap setup. The laser beam (orange lines represent the peripheral rays) is first expanded using a beam expander (BE) consisting of two best-form lenses and then passed through two afocal systems, both composed of two lenses, for distortion-free movement of the trap in three dimensions. After being reflected by a dichroic mirror (DM), the beam is focused by an objective to form the trap at the focal plane. The lenses and mirrors are denoted by L and M, respectively. Microscope illumination (cyan lines) is focused by the condenser and passed through the DM to fall on the CCD camera for imaging of the sample. The inset shows a magnified side view of the beam profile, along with three different position-orientations of a JP in the optical trap and the corresponding optical and thermophoretic forces, as shown in Figure 2.}
\label{fig:setup}
\end{figure*}

A diode-pumped solid-state continuous wave Nd:YAG laser of wavelength $\lambda$ = \SI{1064}{\nm} (Opus 5000, Laser Quantum Ltd.) was used to setup a conventional optical tweezer around an inverted microscope (Nilon Ti2-U) \cite{halder2025interplay,halderOpticalMicromanipulationSoft2024a}. The $\mathrm{TEM}_{00}$ spatial mode linearly polarized beam was expanded and passed through lenses and mirrors for beam steering, before finally focusing it through a 60$\times$ oil-immersion objective with numerical aperture (NA) 1.49 (Nikon CFI Apochromat TIRF 60$\times$) and high transmission at IR ($\approx$ 60\%) to form the optical trap at the focal point, as shown in Figure \ref{fig:setup}. The symmetry of the optical trap was verified by almost identical Gaussian position distributions of a trapped isotropic dielectric microsphere along orthogonal directions, and the trap stiffness was observed to increase linearly with the laser power in the operating range \cite{halderOpticalMicromanipulationSoft2024a}.

\section{Experimental details}

\subsection{Synthesis of Pt-silica JP}\label{janus}

Pt-silica JPs were synthesized following a well-established protocol based on metal deposition on a monolayer of dielectric particles \cite{waltherJanusParticles2008, jiangActiveMotionJanus2010}. In this case, an aqueous dispersion of silica particles with a diameter of $\SI{1.76}{\micro\meter}$ was drop-casted onto a plasma-cleaned glass slide and kept at low temperature (\qty{4}{\degreeCelsius}) overnight for slow evaporation of the solvent. This resulted in a large monolayer of silica particles with hexagonal packing on the glass surface, which was kept in a desiccator for drying. A thin ($\approx$ \SI{5.5}{\nm}) layer of Pt was then deposited onto the monolayer of silica particles in a plasma sputter coater operating in thickness-controlled mode (Figure \ref{fig:TP}(a)). The half-Pt-coated silica JPs were removed from the glass slide by gentle scratching and dispersed in deionized water, followed by multiple washing cycles with SDS to remove the Pt residue. We performed four washing cycles with 1\% SDS solution to remove the Pt residue and an additional three washing cycles in deionized water to remove the SDS surfactant, centrifugation at 3000 rpm for 10 minutes followed by decantation, and resuspension constituting each washing cycle. A diluted suspension of Pt-silica JPs in deionized water was used for the experiments.

\subsection{Fluorescent thermometry}\label{thermometry}

\begin{figure*}[h]
	\centering
	\includegraphics[width=\linewidth]{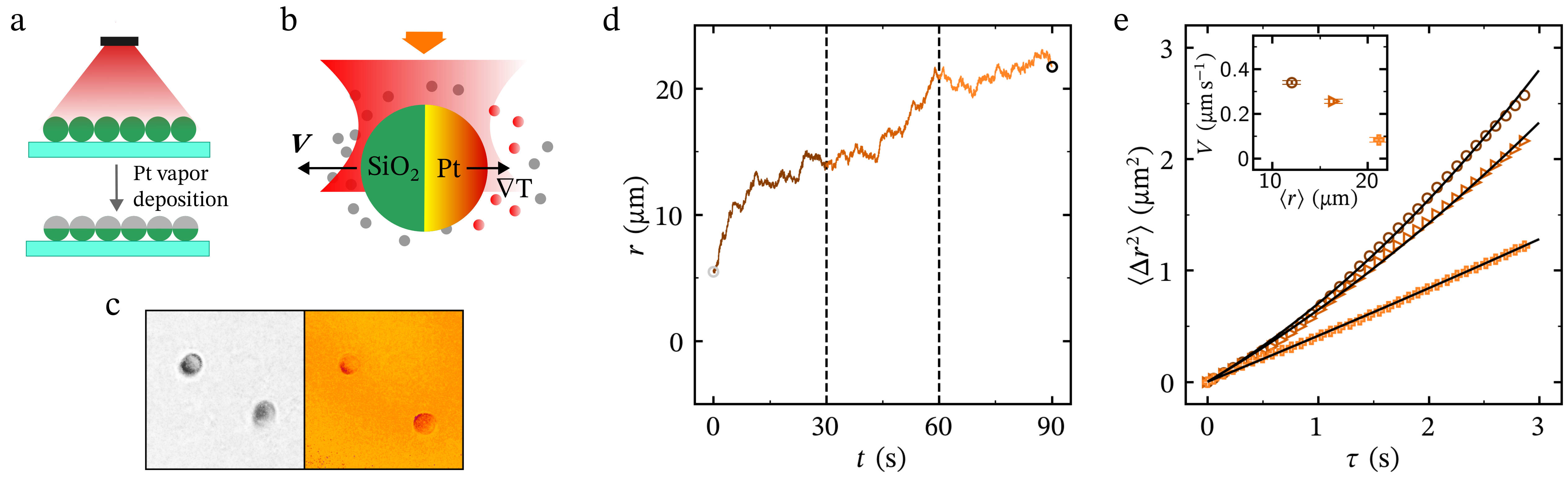}
	\caption{Synthesis, thermophoretic activity, and fluorescence thermometry of Pt-silica JPs. (a) Plasma sputter coating (red) of Pt on a monolayer of silica microspheres (green spheres) to obtain Pt-silica JPs (grey-green spheres) is shown in a schematic (not to scale). (b) Thermophoresis of the Pt-silica JP under laser exposure (red) due to photonic heating on the Pt-coated side (red-yellow gradient), and the resultant propulsion ($\mathbfit{V}$) directed opposite to the temperature gradient ($\mathbfit{\nabla} T$) are exhibited schematically. (c) The brightfield (left) and fluorescence (right) images of two immobilized Pt-silica JPs under laser exposure demonstrate increased temperature on the Pt-coated side (appearing darker in the brightfield image) by reduced emission of Rhodamine B. (d, e) Radial variation in the thermophoretic activity of a Pt-silica JP is shown by (d) a plot of radial position $r (t)$ with time $t$, where the orange gradient (dark to light) denotes the progression of time. (e) MSDs from three consecutive \SI{30}{\s} intervals are shown with color-coded open symbols, while the superimposed black lines represent fitting to Eq. 2. The decrease in propulsion speed $V$, obtained from the fitting, with the mean radial distance $\left\langle r \right\rangle $, is evident from the plot in the inset.}
	\label{fig:TP}
\end{figure*}

We used the temperature-sensitive emission of Rhodamine B to map the temperature difference between the Pt-coated and uncoated sides of the Pt-silica JPs. A 532 nm laser was used to excite Rhodamine B \cite{Ross2001, Natrajan2008, Zhou2019}. Fluorescence thermometry was performed on two different samples. First, a monolayer of silica particles was partially coated with Pt by masking the other part, and was used to check the temperature difference between the Pt-coated and uncoated regions under laser exposure. The significantly reduced emission of Rhodamine B at the Pt-coated part of the monolayer validated the increased temperature (Figure 1(e)). In another sample, two immobilized Pt-silica JP were used, where a similar reduction of the Rhodamine B emission at the Pt-coated side under laser exposure confirmed its higher temperature due to photonic heating (Figure \ref{fig:TP}(c)). This temperature change across a JP in aqueous solution creates a local gradient of approximately a few Kelvin, inducing self-thermophoretic propulsion toward the cooler side \cite{jiangActiveMotionJanus2010} (Figure \ref{fig:TP}(b)).

\subsection{Variation in thermophoretic force with radial distance}
The strength of the thermophoretic force $F_{\mathrm{t}}$, and consequently, the propulsion speed $V$, varies proportionately with the laser power $P$, after a threshold value of $P$, as shown in the inset of Figure 1(g). The laser power also falls radially, following its Gaussian intensity profile. Hence, we characterized the variation in the thermophoretic propulsion speed $V$ with the mean radial distance $\left\langle r \right\rangle $. To this end, we split a typical trajectory exhibiting the net radially outward thermophoretic motion of a JP into three consecutive parts, each of duration \SI{30}{\s} (Figure \ref{fig:TP}(d)). MSDs from the sub-trajectories were plotted and fitted to Eq. 2 to obtain the average value of $V$ for the trajectory segment (Figure \ref{fig:TP}(e)) \cite{Bailey2022}. The variation in $V$ with the respective average radial distance of the trajectory segment $\left\langle r \right\rangle $ is shown in the inset, validating the radial decrease in $V$, with the corresponding variation in the force obtained as $F_{\mathrm{t}} = 6 \pi \eta a V$. Notably, the variation in $V$ with $\left\langle r \right\rangle $ is not strong enough to have any significant influence on the thermophoretic propulsion of a Pt-silica JP when its radial position varies on a small scale, such as in the case of localized and delocalized confinement in an optical trap. In contrast, $F_{\mathrm{t}}$, and consequently $V$, has a stronger dependence on the overall laser power $P$, as shown in the inset of Figure 1(g).

\subsection{Position and orientation tracking}

The time series of the position and orientation of the JPs were obtained from the recorded image frame sequences, which were processed using ImageJ (Fiji) \cite{Schindelin2012}. First, an edge detection filter was applied to separate the particle contour from the background. Subsequently, a Gaussian blur filter was applied to smoothen the bright spots at the centers of the particles. The particle positions, and consequently the trajectories ($x (t)$, $y (t)$), were obtained from the processed sequence of frames using the TrackMate plugin in ImageJ \cite{Ershov2022}. We also tracked the orientation of the JPs ($\hat{n} (t)$), which is defined as the direction perpendicular to the coating interface in the 2D plane and towards the coated side (Figure 1(b)). For this purpose, we first applied a Gaussian blur filter and then identified the uncoated and Pt-coated hemispheres of the JP as brighter and darker sides, respectively. The centers of the dark and bright parts were tracked using TrackMate and the orientation unit vector ($\hat{n}$) was obtained as the line connecting the center of the bright part to that of the dark region. The orientation angle $\phi$ is defined as the angle between $\hat{V} \equiv - \hat{n}$ and $\hat{x}$, as shown in the inset of Figure 3 \cite{Zheng2015}.

\subsection{Variation in trap stiffness with laser power}

In the case of isotropic microspheres, the optical trap stiffness $k$ increases linearly with laser power $P$. However, the interaction of Pt-silica JP with a converging laser beam is more complex because of their anisotropic optical properties. Hence, it remains to be verified whether $k$ increases with $P$ in the case of JPs.

\begin{figure}[bh!]
	\centering
	\includegraphics[width=\linewidth]{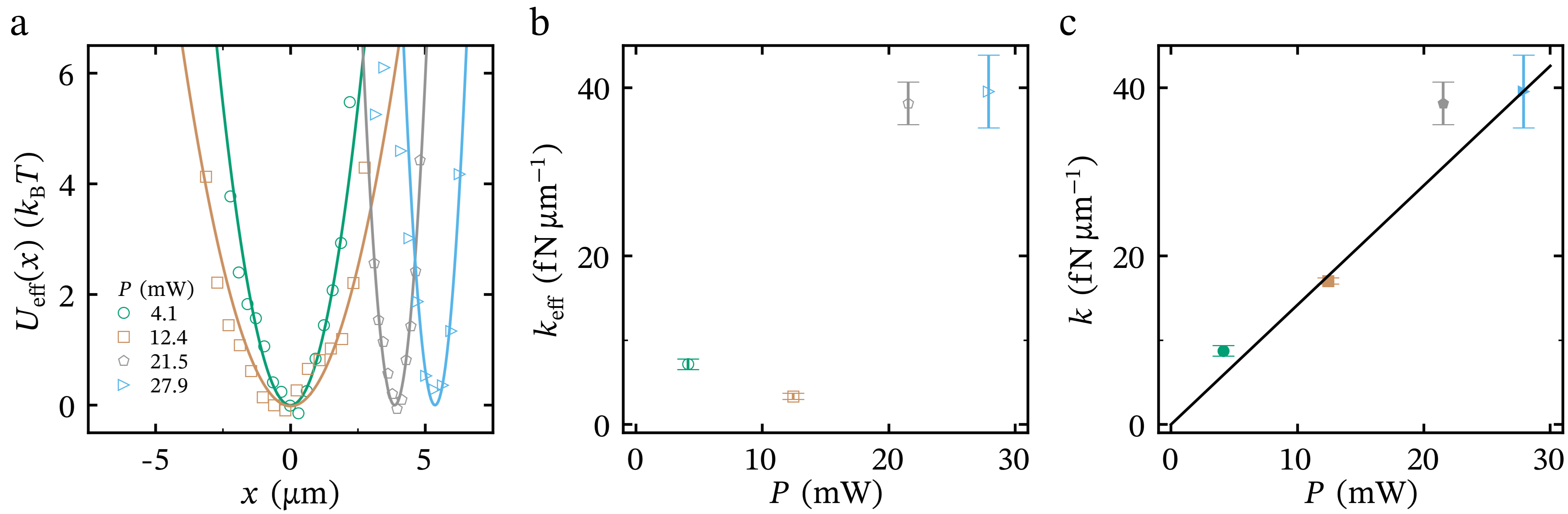}
	\caption{Variation in the effective confinement $U_{\mathrm{eff}}$, corresponding effective trap stiffness $k_{\mathrm{eff}}$, and $k$ with laser power $P$. (a) $U_{\mathrm{eff}} (x)$ (color-coded symbols) at varied $P$ (values given in the inset) are obtained from the corresponding long-time position distributions $p(x)$ (Figure 2(b) and (e)), and are fitted to the quadratic function $k_{\mathrm{eff}} x^2/2$ (color-coded solid lines). (b) The fitting parameter $k_{\mathrm{eff}}$ is plotted against $P$. (c) The trap stiffness $k$, which is derived from $k_{\mathrm{eff}}$, exhibits a linear dependence on $P$, where the linear fit is shown by a black line. Error bars in (b) and (c) represent the fitting uncertainties.}
	\label{fig:k-vs-p}
\end{figure}

To study the variation of trap stiffness $k$ with laser power $P$, we derived the time-averaged effective one-dimensional confining potentials $U_{\mathrm{eff}} (x)$ through Boltzmann inversion of the long-term position distributions of a Pt-silica JP at varied laser powers, as shown in Figure 2(b and e). Thus, we obtained $U_{\mathrm{eff}} (x) = - k_{\text{B}} T \ln p(x)$, where $p(x)$ represents the one-dimensional position distribution (Fig. 2(c and f)). For localized confinement (Figure 2(b)), $p (x)$ is derived from the entire trajectory, whereas $p (x)$ is obtained from a narrow strip on the $x$-axis in the case of delocalized confinement at higher laser powers (Figure 2(e)). All four $U_{\mathrm{eff}}$ plots exhibit excellent fitting with the quadratic function $k_{\mathrm{eff}} x^2/2$ (Figure \ref{fig:k-vs-p}(a)), indicating that the effective confinements are harmonic and providing their effective trap stiffness $k_{\mathrm{eff}}$. The variation in $k_{\mathrm{eff}}$ with $P$ is shown in Figure \ref{fig:k-vs-p}(b). In the localized trapping states at lower laser powers, $k_{\mathrm{eff}}$ changes with propulsion speed $V$ as $k_{\mathrm{eff}} = k / \left(1 + \mathrm{Pe}^2/2\right)$, where $\mathrm{Pe}$ is the P\'eclet number given by $\mathrm{Pe} = V/\sqrt{D_{\mathrm{R}}D_{\mathrm{T}}}$, and $D_{\mathrm{R}}$ and $D_{\mathrm{T}}$ are the rotational and translational diffusion coefficients, respectively \cite{halder2025interplay}. Thus, we derived $k$ from the $k_{\mathrm{eff}}$ values for the localized trapping states, as $V$, $D_{\mathrm{R}}$, and $D_{\mathrm{T}}$ were conveniently obtained for each case by fitting the respective positional autocorrelation functions or MSDs. In the case of delocalized confinement at higher laser powers, $k_{\mathrm{eff}}$ is independent of $V$ and is given by $k_{\mathrm{eff}} = k$ \cite{halder2025interplay}. The $k$ values obtained from $k_{\mathrm{eff}}$, corresponding to two localized and two delocalized trapping states, exhibit a linear increase with the laser power $P$ (Figure \ref{fig:k-vs-p}(c)).

\subsection{Effect of laser beam polarization on position distribution}

\begin{figure}[h!]
	\centering
	\includegraphics[width=\linewidth]{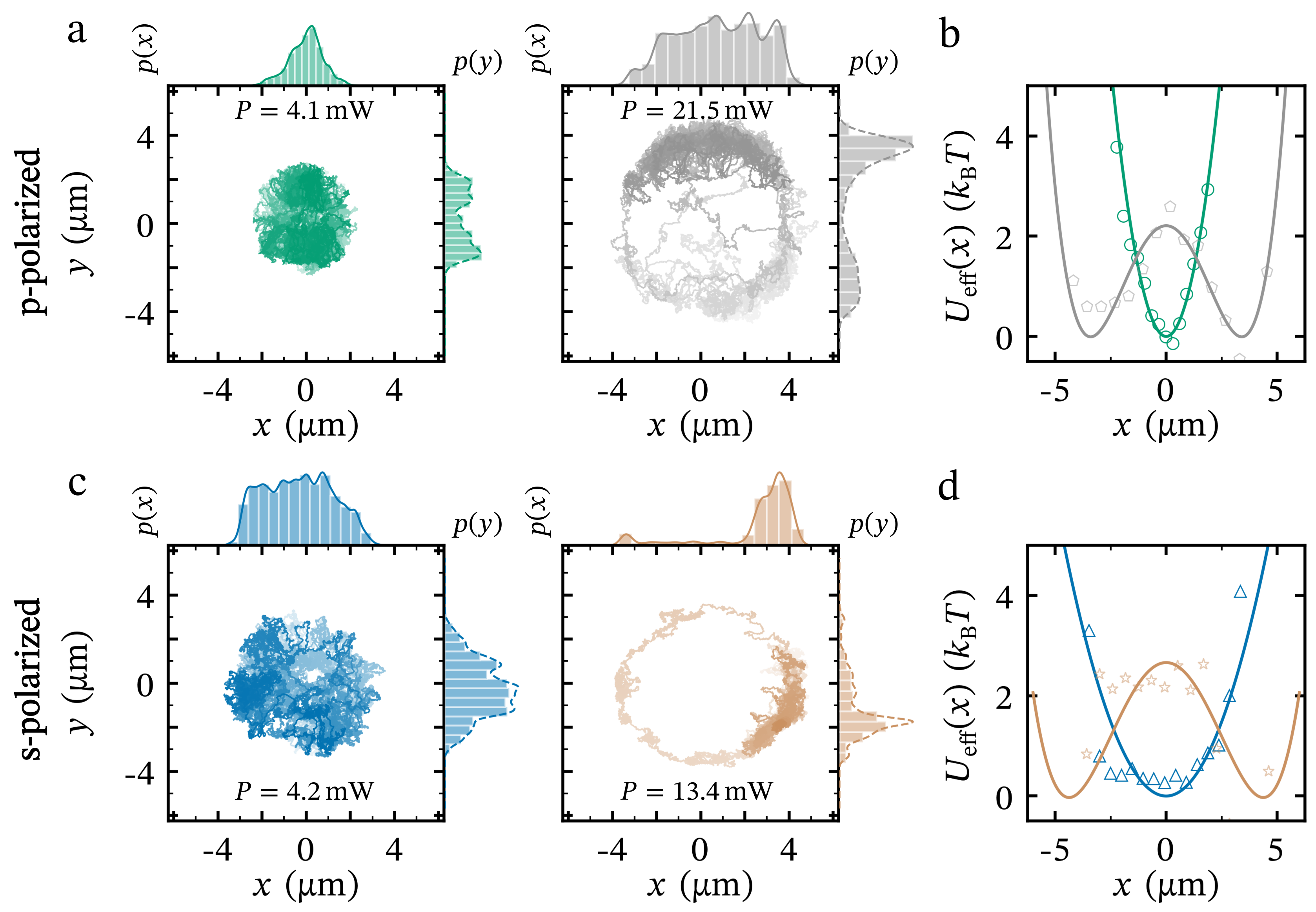}
	\caption{Effect of trapping laser beam polarization on long-time position distribution of optically trapped Pt-silica JPs. (a, c) Position distributions of optically trapped JPs at lower (4.1 and \SI{4.2}{\mW}) and higher (21.5 and \SI{13.4}{\mW}) laser powers, exhibiting localized and delocalized trapping states, are shown for p-polarized (top row, same as those in Fig. 2) and s-polarized (bottom row) laser beams. (b, d) Corresponding effective confining potentials $U_{\text{eff}}$ are shown with color-coded symbols. The solid lines of the same colors represent the fitting to quadratic (for localized trapping states, green and blue) and quartic (for delocalized trapping states, grey and brown) functions. To consistently illustrate the effect of polarization, $U_{\text{eff}}$ for $P$ = \SI{21.5}{\mW} has been derived by considering $p(y)$.}
	\label{fig:polarization}
\end{figure}

We also studied the effect of polarization on the confinement of an optically trapped Pt-silica JP to understand the azimuthal asymmetry in the long-time position distribution, which is more pronounced in the delocalized trapping states (Figure 2(b and e)). Therefore, we repeated our experiments for two orthogonal polarizations, with p-polarized and s-polarized trapping laser beams. We observed that changing from a p-polarized to an s-polarized beam shifted the azimuthal asymmetry in the position distribution to an orthogonal direction. A higher positional probability that is observed along the $y$-axis for the p-polarized trapping laser beam is transformed to become aligned along the $x$-axis when the trapping laser beam is s-polarized. The laser beam polarization does not alter the emergence of localized and delocalized trapping states at lower and higher laser powers, respectively. The effect of polarization is demonstrated in Figure \ref{fig:polarization} by showing the position distributions at lower and higher laser powers, exhibiting localized and delocalized confinements, respectively, and the corresponding effective confining potentials, obtained through Boltzmann inversion, for the p-polarized (top row) and s-polarized (bottom row) trapping laser beams.

We hypothesize that the orientations of the JPs, for which the interface of the coated hemisphere is aligned with the plane of polarization, are energetically favored \cite{Gao2022}. Therefore, certain orientations ($\phi$) are more probable, depending on the plane of polarization of the laser beam. Consequently, an uneven probability distribution in $\phi$ induces a distortion in the azimuthal ($\theta$) position distribution of the optically trapped JPs. Therefore, the distortion in the azimuthal position distribution and its shift to the perpendicular direction upon the orthogonal rotation of the plane of polarization is less apparent in the localized trapping state and more pronounced in the delocalized trapping state, where $\phi$ and $\theta$ are correlated.

\section{Position autocorrelation}\label{corr}

The position autocorrelation of an optically trapped thermophoretically active JP can be obtained by modeling it as an HBABP (Eq. 3), as follows \cite{TenHagen2011, halder2025interplay}: 
\begin{equation}
	\begin{split}
		& \qquad \langle x(t) x(0) \rangle \\
     & = \int_{-\infty}^{0}dt_1 \int_{-\infty}^{t}dt_2 e^{-(t-t_1-t_2)/\tau_k} \langle \xi(t_1)\xi(t_2) \rangle  + V^2 \int_{-\infty}^{0}dt_1 \int_{-\infty}^{t}dt_2 e^{-(t-t_1-t_2)/\tau_k} \langle \cos(\phi(t_1)) \cos(\phi(t_2)) \rangle\\
    & = \int_{-\infty}^{0}dt_1 \int_{-\infty}^{t}dt_2 e^{-(t-t_1-t_2)/\tau_k} \langle \xi(t_1)\xi(t_2) \rangle + V^2 \int_{-\infty}^{0}dt_1 \int_{-\infty}^{t_1}dt_2 e^{-(t-t_1-t_2)/\tau_k} \langle \cos(\phi(t_1)) \cos(\phi(t_2)) \rangle_{t_2<t_1}\\
    & \qquad + V^2 \int_{-\infty}^{0}dt_1 \int_{t_1}^{t}dt_1 e^{-(t-t_1-t_2)/\tau_k} \langle \cos(\phi(t_1)) \cos(\phi(t_2)) \rangle_{t_2>t_1}\\
    & = \int_{-\infty}^{0}dt_1 \int_{-\infty}^{t}dt_2 e^{-(t-t_1-t_2)/\tau_k} 2D_{\text{T}}\delta (t_1-t_2) + \frac{V^2}{2} \int_{-\infty}^{0}dt_1 \int_{-\infty}^{t_1}dt_2 e^{-(t-t_1-t_2)/\tau_k} e^{-(t_1-t_2)/\tau_{\text{R}}} \\
    & \qquad + \frac{V^2}{2} \int_{-\infty}^{0}dt_1 \int_{t_1}^{t}dt_2 e^{-(t-t_1-t_2)/\tau_k} e^{-(t_2-t_1)/\tau_{\text{R}}} + \frac{V^2\cos2\phi_0}{2} \int_{-\infty}^{0}dt_1 \int_{-\infty}^{t_1} dt_2 e^{-(t-t_1-t_2)/\tau_k} e^{-(t_1-t_2)/\tau_{\text{R}}} e^{-4t_2/\tau_{\text{R}}} \\
    & \qquad + \frac{V^2\cos2\phi_0}{2} \int_{-\infty}^{0}dt_1 \int_{t_1}^{t} dt_2 e^{-(t-t_1-t_2)/\tau_k} e^{-(t_2-t_1)/\tau_{\text{R}}} e^{-4t_1/\tau_{\text{R}}} \\
    & = \frac{k_{\text{B}} T}{k}e^{-t/\tau_k} + \frac{V^2}{4} \frac{\tau^2_k\tau_{\text{R}}}{\tau_{\text{R}} + \tau_k}e^{-t/\tau_k} - \frac{V^2}{4} \frac{\tau^2_k\tau_{\text{R}}}{\tau_{\text{R}} - \tau_k}e^{-t/\tau_k} + \frac{V^2}{2} \frac{\tau^2_k\tau^2_{\text{R}}}{(\tau_{\text{R}} - \tau_k)(\tau_{\text{R}} + \tau_k)}e^{-t/\tau_{\text{R}}} \\
    & \quad + \frac{V^2 \cos2\phi_0}{2} \frac{\tau^2_k\tau^2_{\text{R}}}{(\tau_{\text{R}} - 3\tau_k)(2\tau_{\text{R}} - 4\tau_k)} e^{-t/\tau_k} + \frac{V^2 \cos2\phi_0}{2} \frac{\tau^2_k\tau^2_{\text{R}}}{(\tau_{\text{R}} - \tau_k)} \left[ \frac{1}{(\tau_{\text{R}} - 3\tau_k)} e^{-t/\tau_{\text{R}}} - \frac{1}{(2\tau_{\text{R}} - 4\tau_k)}  e^{-t/\tau_k}\right] \\
    & = \frac{k_{\text{B}} T}{k}e^{-t/\tau_k} + \frac{V^2}{2} \frac{\tau^2_k\tau_{\text{R}}}{(\tau_{\text{R}} + \tau_k)(\tau_{\text{R}} - \tau_k)} \left[\tau_k e^{-t/\tau_k} - \tau_{\text{R}}e^{-t/\tau_k}\right] \\
    & \quad + \frac{V^2 \cos2\phi_0}{2} \frac{\tau^2_k\tau^2_{\text{R}}}{(\tau_{\text{R}} - \tau_k)(\tau_{\text{R}} - 3\tau_k)} \left[ \frac{\tau_k}{(\tau_{\text{R}} - 2\tau_k)} e^{-t/\tau_k} +  e^{-t/\tau_{\text{R}}} \right] 
    \end{split}
    \label{eq:corr}
\end{equation}

\section{Numerical simulation}

We simulated the dynamics of an optically trapped phoretically active JP at various laser powers by considering it as an HBABP \cite{halder2025interplay}, which is defined by the Langevin equations given in Eq. 3. Following our observations, we varied the stiffness of the confinement $k$ and the magnitude of the thermophoretic force $F_{\mathrm{t}}$, and hence the propulsion speed $V = F_{\mathrm{t}} / 6 \pi \eta a$ proportionally (Figure \ref{fig:k-vs-p}(c), 1(g)). The insignificant variation of $V$ with the instantaneous radial position of an optically trapped JP, which has been observed to remain confined within a small region around the center of the trap or in an annular region with a narrow radial width, was neglected in the simulations. Therefore, in our simulations, we varied the equilibration time in the harmonic potential, $\tau_k = 6 \pi \eta a / k$, inversely with propulsion speed $V$. The values of all relevant parameters are mentioned in the respective figures.  

The simulations were performed at a time-step of \SI{0.002}{\s} for $10^6$ time-steps to study the steady-state properties of the system. The direction of the propulsion was considered to follow the orientational diffusion of the JP, defined by its orientational diffusion coefficient $D_{\mathrm{R}}$, or an associated characteristic timescale $\tau_{\mathrm{R}} = 1/ D_{\mathrm{R}}$, which depends on the size of the JP and was kept constant ($\tau_{\mathrm{R}}$ = \SI{1}{\s}) for all cases. All other parameters related to the system and JP, such as viscosity, ($\eta$ = \SI{1E-3}{\pascal\cdot\second}), Stokes radius of JP ($a$), and translational diffusion coefficient ($D_{\mathrm{T}}$), were kept unchanged and consistent with each other.

\begin{figure}[ht]
	\centering
	\includegraphics[width=0.98\linewidth]{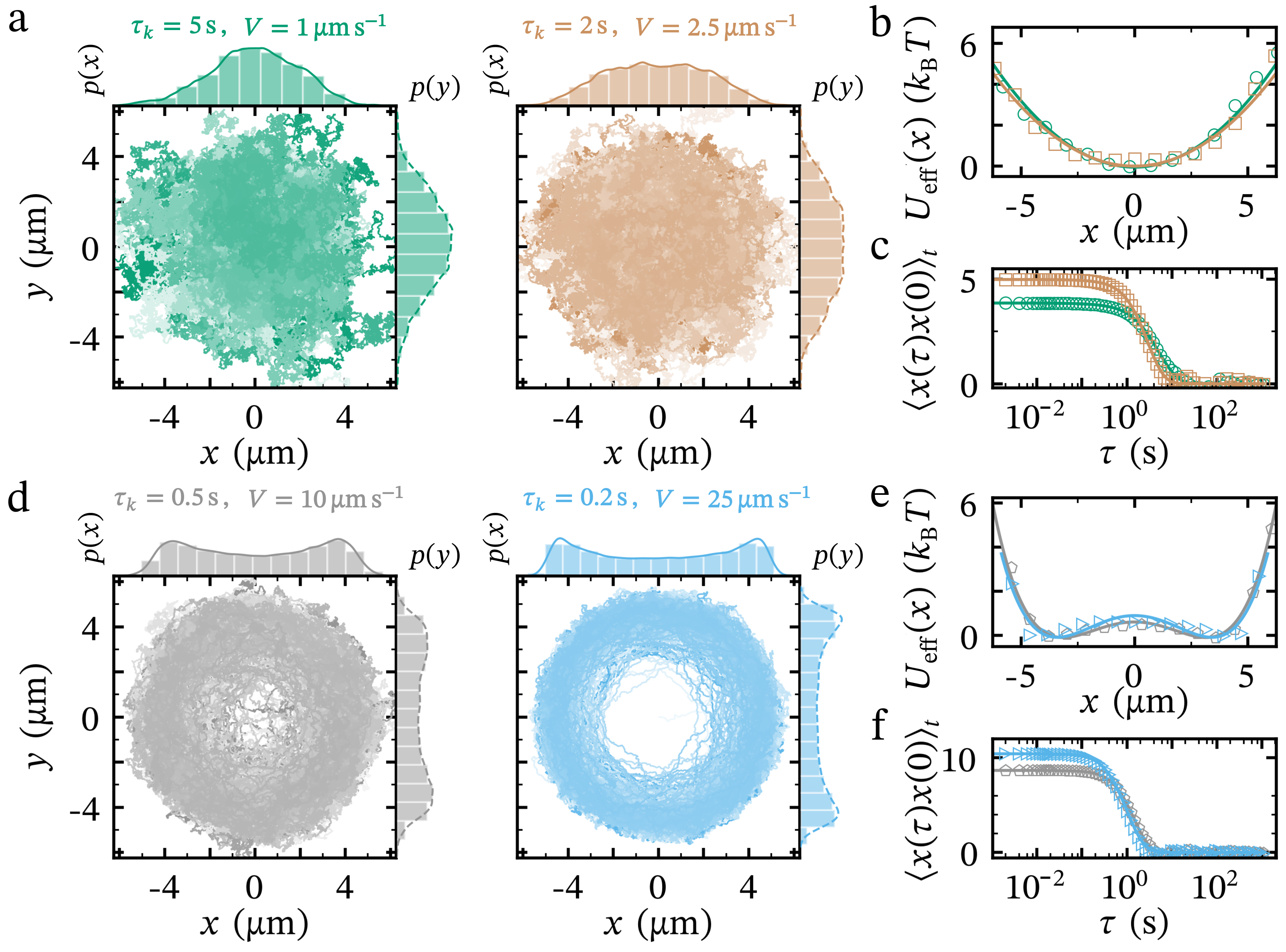}
	\caption{Simulated dynamics of an HBBP representing optically trapped phoretically active JP. (a, d) Simulated trajectories, along with the position distributions, are shown for two longer $\tau_k$ and smaller $V$ values, corresponding to lower laser powers, and two shorter $\tau_k$ and larger $V$ values, signifying higher laser powers. (b, e) Effective confining potentials are plotted with open symbols, whereas the corresponding (b) quadratic and (e) quartic fits are shown with solid lines of the same colors as those of the trajectories. (c, f) Position autocorrelations and fitting with Eq. \ref{eq:corr} are exhibited with color-coded open symbols and solid lines, respectively.}
	\label{fig:SimResults}
\end{figure}

\subsection{Localized and delocalized confinements}

The results from the simulation exhibit characteristic features more apparently because of better statistics from substantially longer simulated trajectories compared to those captured experimentally. The trajectories and corresponding steady-state position distributions (Figure \ref{fig:SimResults}(a, d)) show the exact trends observed in our experiments (Figure 2(b, e)). Localized confinements with Boltzmann-like position distributions at lower laser powers (longer $\tau_k$ and smaller $V$) transform into delocalized annularly confined dynamics with bimodal distributions at higher laser powers (shorter $\tau_k$ and larger $V$). In both cases, the spread of the trajectories increases with the laser power. The effective confining potentials (Figure \ref{fig:SimResults}(b, e)) also exhibit the similar trends as observed from the experimental data (Figure 2(c, f)). At a lower laser power, the effective potential remains harmonic, and the effective stiffness ($k_{\mathrm{eff}}$) decreases with increasing laser power. In contrast, at higher laser powers, the effective potentials fit well with quartic functions, where the minima shift farther apart with increasing laser power. Furthermore, the position autocorrelations in all cases fit well with the analytical prediction given by Eq. \ref{eq:corr} (Figure \ref{fig:SimResults}(c, f)). The position autocorrelation decay times are longer at lower laser powers, indicating a weaker effective potential, compared to those at higher laser powers. This is in good agreement with the reported trend of $k_{\mathrm{eff}}$ \cite{halder2025interplay}.

\subsection{Spin-orbit coupling}

\begin{figure}[ht]
	\centering
	\includegraphics[width=0.7\linewidth]{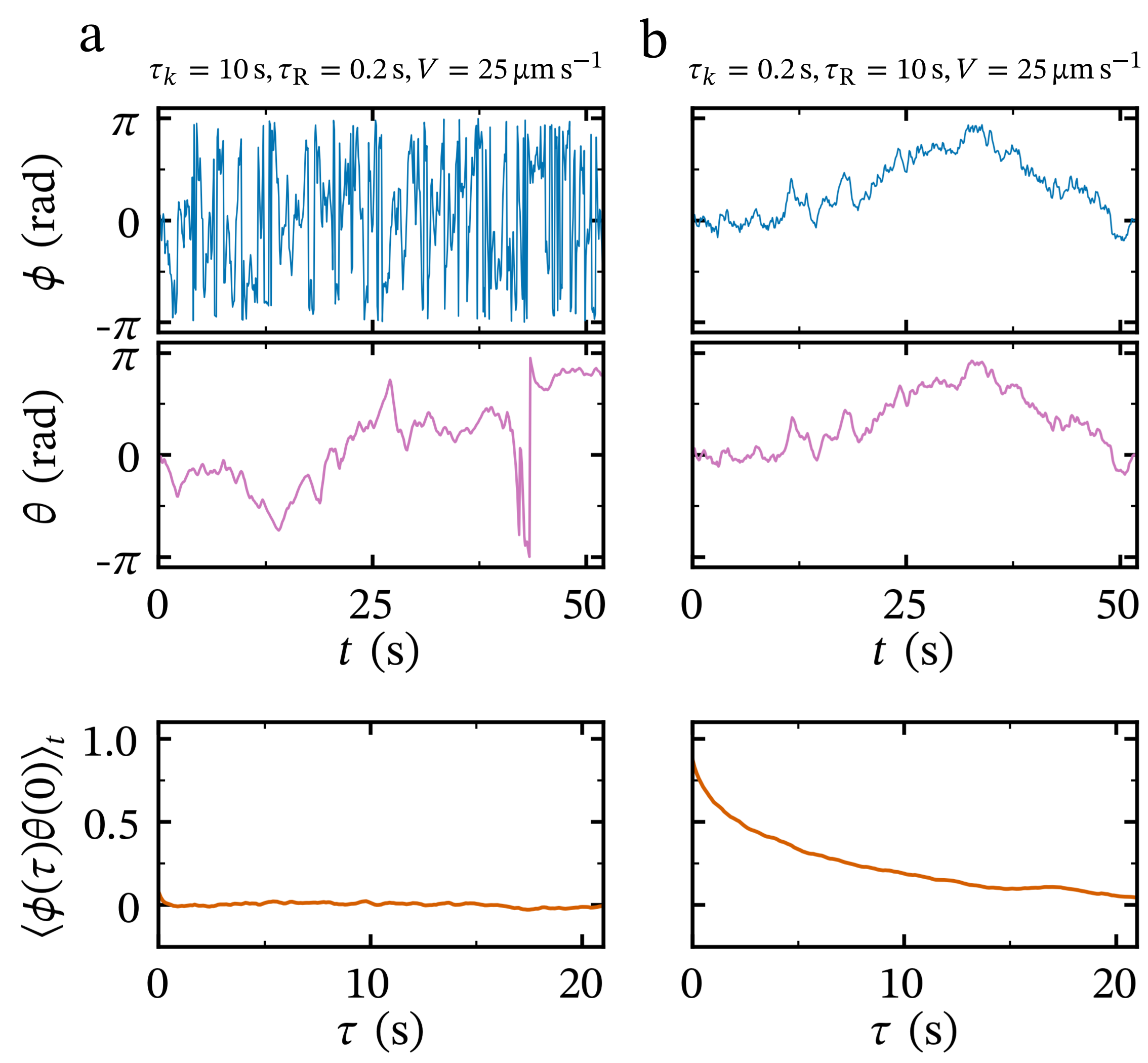}
	\caption{Coupling between the orientational ($\phi (t)$) and orbital ($\theta (t)$) dynamics from the simulation of an active JP in a harmonic well corresponding to localized (a) and delocalized (b) trapping states. Short time series of $\phi (t)$ and $\theta (t)$ are shown along with their cross-correlations in both cases, similar to those in Figure 3. The values of the relevant parameters, \textit{i.e.}, the equilibration time ($\tau_k$), orientational diffusion timescale ($\tau_{\mathrm{R}}$), and propulsion speed ($V$), are given at the top.}
	\label{fig:Coupling-Sim}
\end{figure}

We also calculated the correlation between the instantaneous orientation ($\phi$) and orbital position ($\theta$) from the simulated dynamics to understand the origin of the strong stochastic coupling between them in the delocalized trapping state. The results are similar to those from the experimental observations; a strong and slowly decaying correlation between $\phi$ and $\theta$ persists beyond the orientational diffusion timescale $\tau_{\mathrm{R}}$ when the confinement is delocalized, whereas there is no significant correlation in the localized trapping state (Figure \ref{fig:Coupling-Sim}).

This demonstrates that the emergence of strong coupling is not induced by optical torque, which was not considered in the simulations. Furthermore, the results indicate that the correlation decay-time does not depend on $\tau_{\mathrm{R}}$, but rather on the ratio $\tau_{\mathrm{R}} / \tau_k$, which describes how fast the JP changes its orientation, \textit{i.e.}, the direction of propulsion, in comparison to its equilibration in the harmonic well. At higher laser powers, $\tau_k$ is substantially shorter, and the JP moves away from the center of the trap with a radially outward thermophoretic propulsion $\mathbfit{V}$ and equilibrates at a radial distance before any significant orientational diffusion occurs. A slow change in $\phi$ induces a small azimuthal component in $\mathbfit{V}$ that takes the JP to a new orbital position $\theta$, where $\mathbfit{V}$ again becomes radial (Video S5). Thus, $\theta$ continues to follow a slowly varying $\phi$, which is manifested as a persistent stochastic coupling between them in the delocalized trapping state.

\section{Supporting videos}

{\href{https://drive.google.com/file/d/1OlY6XjiXH8xpVMW_LCGmzbr_DXmnWO6L/view?usp=share_link}{\textbf{Video S1:}}}
Thermophoretic active dynamics of a Pt-silica JP in a defocused laser field. Radially outward dynamics of a thermophoretically active JP with a diameter of \SI{1.76}{\micro\meter} is shown, where the red cross denotes the center of the defocused laser field at a laser power of $P$ = \SI{8.8}{\mW}. As the JP moves away from the center of the laser field, the laser intensity, and hence the thermophoretic activity, decreases, and the dynamics become diffusive-like. The corresponding trajectory is shown in Figure 1(f). The video was recorded at 100 fps with a FLIR Grasshopper monochrome camera attached to a microscope with a 60$\times$ objective. \\

{\href{https://drive.google.com/file/d/1fx_SvoUPz4wT1VoPAkUc1sLg4SosaSw8/view?usp=share_link}{\textbf{Video S2:}}}
Localized optical confinement of a thermophoretically active JP near the focal point. An optically trapped Pt-silica JP remains confined in a local 3D region around the focal point at a lower laser power of $P$ = \SI{4.1}{\mW}. The appearance of the JP changes as its position spontaneously shifts along $\hat{z}$ and crosses the focal plane, staying close to the focal point (red cross). The corresponding trajectory and position distributions are shown in Figure 2(b). The video was recorded at 500 fps with a FLIR Grasshopper monochrome camera attached to a microscope with a 60$\times$ objective. \\

{\href{https://drive.google.com/file/d/1-1pv_RiRXyM_syHzbgXgvG93nLSfgYFG/view?usp=share_link}{\textbf{Video S3:}}}
Delocalized optical confinement of a thermophoretically active JP in an annularly confined region. At a higher laser power of $P$ = \SI{27.9}{\mW}, an optically trapped Pt-silica JP remains away from the focal point (red cross) and exhibits annularly confined dynamics at a lower $z$-plane. The corresponding trajectory and position distributions are shown in Figure 2(e). The video was recorded at 500 fps with a FLIR Grasshopper monochrome camera attached to a microscope with a 60$\times$ objective.\\

{\href{https://drive.google.com/file/d/1aCHjMdPPhdLxt2xTzsJW-rwLBcZyX8FJ/view?usp=share_link}{\textbf{Video S4:}}}
Coexistent dynamically stable optical trapping of two Pt-silica JPs. Two optically trapped JPs with different Pt-coating profiles remain confined at different delocalized annular regions within the 3D laser field at a laser power of $P$ = \SI{12.4}{\mW}. While one JP stays closer to the focal point (red cross), the other remains away from the focal point and at a lower $z$-plane. The corresponding trajectories are shown in Figure 4(b). The video was recorded at 500 fps with a FLIR Grasshopper monochrome camera attached to a microscope with a 60$\times$ objective.\\

{\href{https://drive.google.com/file/d/1pBZP3IznVtL25WicY1C9my4H265Scz5_/view?usp=sharing}{\textbf{Video S5:}}}
Stochastic spin-orbit coupling from the HBABP simulation. A strong correlation between the orientational ($\phi (t)$) and orbital ($\theta (t)$) dynamics is demonstrated by the change in direction of the black and green arrows that denote $\hat{V}$ and $\hat{r}$, which are perpendicular to $\hat{\phi}$ and $\hat{\theta}$, and thus provide a visual quantification of $\phi (t)$ and $\theta (t)$, respectively. This occurs in the delocalized trapping state, where the JP stays away from the center of the harmonic potential, represented by the orange gradient, and exhibits annularly confined dynamics. For this simulation, the parameter values were $\tau_k$ = \SI{0.2}{\s}, $\tau_{\mathrm{R}}$ = \SI{10}{\s}, and $V$ = \SI{25}{\um/\s}. The dynamics was simulated at 500 fps and is presented at a playback speed of 4$\times$.\\

\clearpage

\bibliography{si_references.bib}